\documentclass[12pt,preprint]{aastex}
\usepackage{graphicx}  

\slugcomment{Submitted to {\it The Astrophysical Journal Letters}}

\shorttitle{Sodium and Oxygen Abundances in NGC 6791}
\shortauthors{Cunha et al.}

\begin{document}


\title{Sodium and Oxygen Abundances in the Open Cluster NGC 6791 from APOGEE
H-Band Spectroscopy}


\author{Katia Cunha}
\affil{Observat\'orio Nacional, S\~ao Crist\'ov\~ao, Rio de Janeiro, Brazil and}
\affil{University of Arizona, Tucson, AZ 85719, USA}
\email{kcunha@on.br}

\author{Verne V. Smith}
\affil{National Optical Astronomy Observatories, Tucson,  AZ 85719 USA}

\author{Jennifer A. Johnson}
\affil{Department of Astronomy, The Ohio State University, Columbus, OH  43210}

\author{Maria Bergemann}
\affil{Max-Planck-Institut fur Astronomy, 69117, Heidelberg, Germany and 
Institute of Astronomy, University of Cambridge, CB3 0HA, Cambridge, UK}

\author{Szabolcs M{\'e}sz{\'a}ros}
\affil{ELTE Gothard Astrophysical Observatory, H-9704 Szombathely, 
Hungary and
Department of Astronomy, Indiana University, Bloomington, IN, 47405, USA} 

\author{Matthew D. Shetrone}
\affil{Department of Astronomy and McDonald Observatory, University of Texas, Austin,  TX 78712  USA}

\author{Diogo Souto}
\affil{Observat\'orio Nacional, S\~ao Crist\'ov\~ao, Rio de Janeiro, Brazil}

\author{Carlos Allende Prieto}
\affil{Instituto de Astrof\'{\i}sica de Canarias, 38205 La Laguna, Tenerife, Spain and}
\affil{Universidad de La Laguna, Departamento de Astrof\'{\i}sica, 38206 La Laguna, Tenerife, Spain}

\author{Ricardo P. Schiavon}
\affil{Astrophysics Research Institute, Liverpool John Moores University, Liverpool L3 5RF UK}

\author{Peter Frinchaboy}
\affil{Texas Christian University, Fort Worth, TX, USA}

\author{Gail Zasowski}
\affil{Department of Physics and Astronomy, Johns Hopkins University, Baltimore, MD 21218, USA}

\author{Dmitry Bizyaev}
\affil{Apache Point Observatory, Sunspot, NM 88349 USA and}
\affil{Strenberg Astronomical Institute, Moscow  119992, Russia}

\author{Jon Holtzman}
\affil{Department of Astronomy, New Mexico State University, Las Cruces, NM  88003 USA}

\author{Ana E. Garc{\'\i}a P\'erez and Steven R. Majewski}
\affil{Department of Astronomy, University of Virginia, Charlottesville,  VA  22904 USA}

\author{David Nidever}
\affil{Department of Astronomy, University of Michigan, Ann Arbor, MI 48109, USA}

\author{Timothy Beers}
\affil{Department of Physics and JINA Center for the Evolution of the Elements, University of Notre Dame, Notre Dame, IN 46556, USA}

\author{Ricardo Carrera}
\affil{Instituto de Astrof\'isica de Canarias, La Laguna, 38200 Tenerife, Spain and
Departamento de Astrof\'isica, Universidad de La Laguna, 38200 Tenerife, Spain}

\author{Doug Geisler}
\affil{Departamento de Astronomia, Casilla 160-C, Universidad de Concepcion, Chile}

\author{James Gunn}
\affil{Department of Astrophysics, Princeton University, Princeton, NJ 08540, USA}

\author{Fred Hearty}
\affil{Department of Astronomy and Astrophysics, The Pennsylvania State University, University Park, PA 16802, USA}

\author{Inese Ivans}
\affil{Department of Physics and Astronomy, The University of Utah, Salt Lake City, UT 84112, USA}

\author{Sarah Martell}
\affil{Department of Astrophysics, School of Physics, University of New South Wales, Sydney, NSW 2052, Australia}

\author{Marc Pinsonneault}
\affil{Department of Astronomy, The Ohio State University, Columbus, OH 43210, USA}

\author{Donald P. Schneider}
\affil{Department of Astronomy and Astrophysics, The Pennsylvania State University, University Park, PA 16802, USA and
and Institute for Gravitation and the Cosmos, The Pennsylvania State University, University Park, PA 16802, USA}

\author{Jennifer Sobeck}
\affil{Department of Astronomy, University of Virginia, Charlottesville,  VA  22904 USA}

\author{Dennis Stello}
\affil{Sydney Institute for Astronomy (SIfA), School of Physics, University of Sydney, NSW 2006, Australia and
Stellar Astrophysics Centre, Department of Physics and Astronomy, Aarhus University, DK-8000 Aarhus C, Denmark}

\author{Keivan G. Stassun}
\affil{Department of Physics \& Astronomy, Vanderbilt University, Nashville, TN 37235, USA and
Department of Physics, Fisk University, Nashville, TN 37208, USA}

\author{Michael Skrutskie}
\affil{Department of Astronomy, University of Virginia, Charlottesville,  VA  22904 USA}

\author{John C. Wilson}
\affil{Department of Astronomy, University of Virginia, Charlottesville,  VA  22904 USA}




\begin{abstract}
The open cluster NGC 6791 is among the oldest, most massive and metal-rich open clusters in the Galaxy.
High-resolution $H$-band spectra from the Apache Point Observatory Galactic Evolution
Experiment (APOGEE) of 11 red giants in NGC 6791 are analyzed
for their chemical abundances of iron, oxygen, and sodium. The abundances of these
three elements are found to be homogeneous (with abundance dispersions at the level of $\sim$ 0.05 - 0.07 dex) in these
cluster red giants, which span much of the red-giant branch (T$_{\rm eff}$ $\sim$ 3500K - 4600K),
and include two red-clump giants. From the infrared spectra, this cluster is confirmed to be among 
the most metal-rich clusters in the Galaxy ($<$[Fe/H]$>$ = 0.34 $\pm$ 0.06), and is
found to have a roughly solar value of [O/Fe] and slightly enhanced [Na/Fe]. 
Non-LTE calculations for the studied Na I lines in the APOGEE spectral region 
($\lambda$16373.86\AA\ and $\lambda$16388.85\AA) indicate only small departures from 
LTE ($\leq$ 0.04 dex) for the parameter range and metallicity of the studied stars.  
The previously reported double population of cluster members with different Na abundances 
is not found among the studied sample. 
\end{abstract}



\section{Introduction}

The open cluster NGC 6791 is a notable object, being one of the oldest 
($\sim$ 8 Gyr old; e.g., Harris \& Canterna 1981; King et al. 2005; Brogaard et al. 2012) 
and most metal-rich open clusters in the Galaxy ([Fe/H]$\sim$ +0.4
\footnote{
[X/Q]=(A(X)$_{\rm Star}$ -- A(X)$_{\odot}$) -- (A(Q)$_{\rm Star}$ -- A(Q)$_{\odot}$). 
}
; e.g., Peterson \& Green 1998; Gratton et al. 2006; Carraro et al. 2006; Origlia et al. 2006; Carretta et al. 2007;
Boesgaard et al. 2009).
This cluster's color-magnitude diagram displays a few unusual features that heighten
interest in this object, including a well-defined
binary sequence falling above the main sequence, several alleged blue horizontal-branch
stars (Platais et al. 2011), which is unusual for a metal-rich stellar population (Brogaard et al. 2012), 
along with a white-dwarf cooling sequence that exhibits two luminosity peaks (Bedin et al. 2008a,b).  

Other unusual aspects of NGC 6791 are the color-width of the red-giant branch (RGB; Kinman 1965) and
color variations of the main-sequence turn-off stars (Twarog et al. 2011). These features could
be explained as either extended star formation (over $\sim$ 1 Gyr), a process not identified in any
other open cluster, or, spatial variations in the reddening (Twarog et al. 2011; Platais et al. 2011; 
Brogaard et al. 2012).
A recent result that relates to the possibility of extended star formation in NGC 6791 is
from Geisler et al. (2012), who found two distinct groups differing in their respective
sodium abundances.  These two Na-abundance groups were proposed to be two
separate stellar populations within NGC 6791; this would be the
first time such distinct stellar populations had been identified in an open cluster, although
this is now commonly found in most globular clusters (Gratton et al. 2012). These distinct
populations are responsible for the Na and O abundance anti-correlations which are
the signature of a globular cluster. 
As NGC 6791 is among the most massive open clusters, whether it has anti-correlations in 
the light elements is critical for setting the lower limit in mass where such variations 
appear (e.g. Geisler et al. 2012). In this context, Bragaglia et al. (2014) 
found that the Na abundances in a sample of NGC 6791 members can be described by a 
single Na abundance to within $\sim$ 0.1 dex.

Due to its high metallicity (as well as location within the \emph{Kepler} field of view),
NGC 6791 was targeted as a calibration cluster by the Apache Point Observatory 
Galactic Evolution Experiment (APOGEE; Majewski et al. 2012; Zasowski et al. 2013), 
one of four experiments that are part of the Sloan Digital Sky Survey III (SDSS-III;
Eisenstein et al. 2011). 
APOGEE obtained H-band high-resolution spectra of red giants with the goal of measuring radial velocities 
and chemical abundances of up to 15 elements in over 100,000 stars 
in the Galactic bulge, disk, and halo. 
NGC 6791 was selected for the present analysis due to both its high metallicity, which places it at the high end of
the APOGEE target metallicities, and the possibility that it contains more than one stellar
generation.  In this Letter, abundance results for iron, oxygen, and sodium are presented 
to investigate whether the selected sample reveals two sodium populations.

\section{The APOGEE Data}

APOGEE data are obtained with a cryogenic, bench-mounted,
300-fiber spectrograph
(Wilson et al. 2010)
attached to the Sloan 2.5m telescope (Gunn et al. 2006).
The APOGEE spectra are high-resolution (R=$\lambda$/$\Delta$$\lambda$$\sim$ 22,500) 
and span the wavelength range from $\lambda$15,100\AA\ -
$\lambda$17,000\AA, with spectra recorded on three separate detectors.
Gaps between the detectors result in three spectral regions covering roughly
$\lambda$15,100\AA\ - $\lambda$15,800\AA, $\lambda$15,900\AA\ -
$\lambda$16,400\AA, and $\lambda$16,500\AA\ - $\lambda$16,900\AA.
An automated data processing pipeline for APOGEE (Nidever et al.,
in preparation) produces one-dimensional, wavelength-calibrated,
flux-calibrated spectra.  These spectra have had terrestrial airglow
lines removed via dedicated sky fibers in each field, and
have telluric absorption lines (from H$_{2}$O, CO$_{2}$, and
CH$_{4}$) corrected using measurements from fibers placed on early-type stars in each field.

Twenty-nine red-giant members of NGC 6791 have been observed with APOGEE; 
their membership has been discussed in Frinchaboy et al. (2013).
All of the observed spectra were inspected and had similar S/N at a given magnitude; overall
S/N increased for the brighter targets and was roughly the same (10-20\%) across the 3 detectors. 
Eleven targets with stellar parameters covering the RGB,
as well as the RC, were selected for this study (Table 1).
The selected targets had spectra with S/N from $\sim$90-130 (for the faintest targets) to
$\sim$200-700 (for the brightest targets), depending on their location on the RGB.
Tofflemire et al. (2014) have recently performed a radial-velocity membership
study of NGC 6791 and their results indicate that all targets analyzed
here are cluster members.

\section{Stellar Parameters}

Fundamental stellar parameters for the NGC 6791 targets are presented in Table 1.
Effective temperatures were derived from the dereddened 2MASS ($J-K_{\rm s}$)
colors and the $T_{\rm eff}$ -- color calibrations from both Bessell et al.
(1998) and Gonzalez Hern\'andez \& Bonifacio (2009), with the tabulated
$T_{\rm eff}$ being the average of the two scales. 
The adopted reddening, $E(B-V)$=0.14, is an average from two recent determinations (Brogaard et al. 2012;
Geisler et al. 2012), and translates to $E(J-K_{\rm s}$)=0.07 (Cardelli et al. 1989).
Brogaard et al. (2012) found variable reddening across the cluster with
$\Delta E(J-K) =$0.025. A change in reddening of this amount translates to a maximum 
change in T$_{eff}$ of $\sim$ 50-60K and a corresponding change in the Fe, O and Na 
abundances of 0.02, 0.12, 0.03 dex, respectively; such changes are within the overall
uncertainties and a mean reddening has been adopted.

Surface gravities were derived from the effective temperatures, along with the
red-giant masses and luminosities.  
The luminosities were calculated using the distance modulus found in 
Basu et al. (2011) of ($m-M$)=13.1, with bolometric corrections from Bessell et al. (1998),
and using a cluster red giant-mass of 1.2M$_{\odot}$ (Basu et al. 2011).
Two of the target stars are RC giants, and there is evidence for a small mass
difference between the RGB and RC in NGC 6791 (Miglio et al. 2012),
with M$_{\rm RGB}$ - M$_{\rm RC}$= 0.08M$_{\odot}$.  This small mass difference results
in a difference in log g of 0.03 dex, small enough to be ignored in computing the surface
gravities. It is noted that both the Basu et al. (2011) and Miglio et al. (2012) masses are within 
$\sim$0.05M$_{\odot}$ of RGB mass determinations from eclipsing binaries in NGC 6791, 
M$_{\rm RGB}$=1.15M$_{\odot}$ (Brogaard et al. 2012).

\section{Abundance Analysis}

The model atmospheres adopted here are 1-D static
models computed with Kurucz's ATLAS9 code (Kurucz 1993) for the 
APOGEE Stellar Parameter and
Chemical Abundance Pipeline (ASPCAP) by M\'esz\'aros et al. (2012).
Synthetic spectra were calculated in LTE using the spectrum 
synthesis code MOOG (Sneden 1973) and the APOGEE line list (Shetrone et al., in preparation).
The oscillator strengths for the transitions of CO, OH and CN are from Goorvitch (1994), Goldman et al. (1998), and Kurucz (1993) and 
Melendez \& Barbuy (1999), respectively. 

Iron abundances and microturbulent velocities ($\xi$) were derived using 
the sample of nine Fe I lines selected in Smith et al. (2013). 
Although there are many Fe I lines in the APOGEE region, 
the advantage of using this smaller set of iron lines is that they suffer less from blending with other 
features, thus being to a certain degree independent of the adopted C, N, and O abundances.
In addition, these lines cover a range in line strength, which makes them suitable for estimating the 
microturbulent velocity parameter. Iron abundances were derived for a range
of microturbulent velocities; the adopted microturbulence was the one
that produced the smallest iron abundance scatter, which also corresponds to
near-zero slopes between A(Fe) and line strength.

Oxygen abundances were derived from vibration-rotation lines of OH in five spectral regions: 
$\lambda$$\lambda$ 15277 -– 15282\AA; 15390 -– 15392\AA; 15504 -– 15507\AA; 15568 -– 15573\AA\ 
and 16189 -– 16193\AA\ (Smith et al. 2013). 
A sample spectrum containing one of the selected regions with two OH lines is presented in 
Figure 1 (top panel). The star shown, 2M19211007+3750008, is one of the hotter
giants analyzed (T$_{\rm eff}$=4435K) and was chosen for illustration because the OH lines become weak
in the hotter giants; here the lines reach about 10-15\% in depth.  The small $\sim$1-2\%
deviations from a smooth spectrum are caused by imperfect 
subtraction of telluric emission lines in these same OH lines observed in the star.  The
solid curves are synthetic spectra with three different oxygen abundances, as indicated
in the figure.
The determination of oxygen abundances from OH lines requires that the carbon abundances (derived from CO lines), 
and to a lesser degree the nitrogen abundances (derived from CN lines), are all solved self-consistently and
that the molecular equilibrium is satisfied (e.g., Cunha \& Smith 2006; Ryde et al. 2009). 
The details concerning the CO and CN lines/bandheads 
used in our analysis can be found in Smith et al. (2013); the abundance results 
for C and N, along with a number of other elements for our targets, will be presented in a future 
paper (Cunha et al., in preparation).
 
Sodium abundances were obtained from the two well-defined Na I lines falling in the 
middle APOGEE detector, with this doublet arising from the $2p^6 4p ^2P^o$ level with 
$\chi$$_{\rm low}$=3.753 eV, to the $2p^6 6s ^2S_{1/2}$ level with $\chi$$_{\rm hi}$=4.510 eV. 
The air wavelengths are $\lambda$16373.86\AA\ and $\lambda$16388.85\AA, with values 
of log $gf$ = -1.318 and -1.018, respectively (Shetrone et al. in preparation).  
One of the observed Na I lines is illustrated in Figure 1 (bottom panel; solid circles). 
Synthetic spectra with three different sodium abundances are also displayed as the
solid curves.   

\section{Abundance Results}

The mean Fe, O, and Na abundances, as well as the microturbulent 
velocities adopted for the target stars, are found in Table 1. In most instances, the
line-to-line abundance scatter (standard deviations) is less than 0.1 dex.
The sensitivities of the oxygen and iron abundances to uncertainties 
in the adopted stellar parameters are discussed in our previous study (Smith et al. 2013). 
Smith et al. (2013) did not include sodium, so the abundance sensitivies of 
the two studied Na I lines to stellar parameters are presented here.  
The procedure used is the same as Smith et al. (2013), and involves 
perturbing the stellar parameters of T$_{\rm eff}$, log g, $\xi$, and overall model
metallicity ([m/H]) and determining how much these parameter
perturbations change the derived Na abundances.  The relevant
covariant terms between primary stellar parameters are included
with perturbations of $\Delta$T$_{\rm eff}$= $\pm$ 50K, $\Delta$log g=
$\pm$ 0.2 dex, $\Delta$$\xi$= $\pm$ 0.2 km-s$^{-1}$, $\Delta$[m/H]=
$\pm$ 0.1 dex.  Two representative model atmospheres were perturbed,
with one a 'hot' model (T$_{\rm eff}$=4500K, log g=2.50, $\xi$=1.3
km-s$^{-1}$, [m/H]=+0.4) and the other a 'cool' model (T$_{\rm eff}$=
3530K, log g=0.8, $\xi$= 1.8 km-s$^{-1}$, [m/H]=+0.4).  Given these
two models and perturbations, the respective Na abundance changes are
$\Delta$A(Na)=$\pm$ 0.06 dex for the hot model and $\Delta$A(Na)=$\pm$ 0.09 dex
for the cool model.  These variations are similar to the
abundance differences found between the two individual Na I lines as
listed in Table 1.

Non-LTE abundance corrections for Na lines were computed using
the statistical equilibrium and line formation codes described in Bergemann et al. (2012a, 2012b, 2013)
and the model atom described in Gehren et al. (2004).
In this previous work, the model was applied to the analysis of optical Na I lines in solar-type
stars, and indicated a significant negative correction relative to LTE at low metallicity,
typically [Fe/H] $< -1$.  Here, we extend these calculations to the
H-band Na I lines, and derive non-LTE corrections for the transitions $\lambda$16373.86\AA\  and 16388.85\AA.
To our knowledge, these are the first non-LTE
results for the sodium spectral lines in the H-band; we find that the two
IR Na I lines are largely free from departures from LTE over the T$_{\rm eff}$
and log g ranges covered by the high-metallicity red giants analyzed here.
The individual lines have non-LTE corrections (A(Na)$_{\rm non-LTE}$ - A(Na)$_{\rm LTE}$)
that range from $-0.01$ to $-0.04$ dex; corrections of this
size have no significant impact on our conclusions.

\section{Discussion}

The sample of 11 RGB and RC stars studied in the open cluster NGC 6791 have Fe 
abundances that are quite homogeneous: the mean value is A(Fe)=7.84 $\pm$ 0.06.  
The abundance dispersion is consistent with the expected errors in the derivation of A(Fe). 
The mean iron abundance for the target stars is quite enhanced relative to solar, 
confirming previous results in the literature that   
NGC 6791, although very old, is among the most metal-rich populations in the Galaxy. 
The oxygen and sodium abundances are also found
to be homogeneous, with mean values and standard deviations of A(O)=9.10 $\pm$ 0.06 and A(Na)=6.75 $\pm$ 0.07,
which yield a near-solar [O/Fe]=+0.01 and a slightly enhanced [Na/Fe]=+0.17.

Very few studies in the literature have derived both Na and O abundances for NGC 6791 members. 
As discussed previously, Geisler et al. (2012) 
obtained the unexpected result of two populations in sodium in this open cluster from an analysis 
of optical spectra obtained with Keck/HIRES and WIYN/Hydra.
Figure 2 shows the [Na/Fe] vs. [O/Fe] abundances for our 11 stars (blue circles), 
in comparison with the bracket abundances taken directly from Geisler et al. (2012; red triangles).
The two studies adopted slightly different solar reference values; the solar 
abundances adopted here are: A${_\odot}$(Fe)= 7.50; A${_\odot}$(O)=8.75; A${_\odot}$(Na)= 6.24 (Caffau et al. 2008; Asplund et al. 2009).
If the Geisler et al. (2012) results were put on this scale, their [Na/Fe] and [O/Fe] 
values would increase by +0.08 and +0.05 dex, respectively. 
Although the [O/Fe] abundances in the two studies are in reasonable agreement, 
it is clear that the Na abundance distribution of NGC 6791 members obtained here
does not overlap with the behavior of Na found in the Geisler et al.(2012) sample.
The two studies have six stars in common (the green lines in Figure 2 connect these results)
and the [Na/Fe] abundances for these 6 stars are systematically lower in this study 
relative to Geisler et al., with an average difference of $\Delta$[Na/Fe]=-0.20 dex. 
In addition, Geisler et al. (2012) find a population of stars with lower Na abundances 
that is not matched by the results obtained for our sample. 

As a further comparison, Figure 3 shows [Na/Fe] vs.
T$_{\rm eff}$ for this study and Geisler et al. (2012).   It is not expected that there would be a
dependence of the Na abundance with T$_{\rm eff}$, which maps
stellar evolution along the RGB.  The values of [Na/Fe] derived here exhibit a
small increase with decreasing temperature, but the change is rather small, at
about 0.15 dex over $\sim$1100K.  The two hottest stars
in our sample are clump giants, and have already evolved up the RGB, to
low values of T$_{\rm eff}$, and then moved to the higher-temperature RC
after having undergone the He-core flash. In contrast, the
results by Geisler et al. (2012) exhibit a discontinuity in [Na/Fe] abundance 
at around T$_{\rm eff}$$\sim$4500K.  The NGC 6791 members from Geisler
et al. (2012) that have elevated values of [Na/Fe], representing the Na-rich population,
are only found for red giants with T$_{\rm eff}$$\le$  4500K, with the lower-Na
abundance giants falling at higher T$_{\rm eff}$'s. 
Geisler et al. (2012) argue, however, that the most direct evidence for a Na-spread in their sample
is among the RC stars, where half of their sample of RC stars showed significant
Na enhancement.

The Na and O abundances obtained here, and their respective abundance dispersions, agree
well with those of the recent study by Bragaglia et al. (2014), who analyzed optical
spectra of 35 stars in NGC 6791 from Keck/HIRES and WIYN/Hydra. 
Bragaglia et al. (2014) find average abundances (and abundance dispersions) of 
A(Fe)=7.87 $\pm$ 0.06 ($\Delta$(This study - Bragaglia)= -0.03),
A(O)=9.00 $\pm$0.09 ($\Delta$(This study - Bragaglia)= +0.10) and 
A(Na)= 6.70 $\pm$ 0.10 ($\Delta$(This study - Bragaglia)=+0.05). 
The comparison between the results from their optical and our H-band abundance analyses are very good, with
differences of less than 0.1 dex and similar dispersions ($\sigma$ $\sim$ 0.06 - 0.10 dex).
Earlier results from the optical for two RC stars by Carretta et al. (2007) were much lower
(they found [O/Fe]=-0.35 for the two stars analyzed). The Na abundances for their two 
studied stars differed by 0.3 dex ([Na/Fe]=+0.28 and -0.02); possibly indicating
an intrinsic abundance scatter for Na, but this scatter is not confirmed in the much 
larger sample by Bragaglia et al. (2014).

Within the sample of 11 stars analyzed in this study, we do not find evidence of two 
distinct populations in sodium in NGC 6791. 
On the contrary, our results are well represented by a single population, with 
an abundance scatter of the order of the expected abundance analysis errors.
We note that we have not analyzed any of the stars that Geisler et al. (2012) found to 
be Na-poor. It is not impossible that only stars from one population, from an underlying 
two-population distribution, were selected for this study, as well as in the sample of
Bragaglia et al. (2014). However, this is unlikely given that roughly 40 percent of 
the stars in Geisler et al. (2012) were from the low-Na population, while 60 percent from the high-Na population.



\section{Conclusions}

Based on $H$-band high-resolution APOGEE spectra of 11 red-giant members of NGC 6791, 
this old open cluster is found to be quite metal rich and chemically homogeneous,
with a mean and standard deviation of $<$[Fe/H]$>$=+0.34$ \pm$ 0.06.  
We also find that both oxygen and sodium are enhanced along with iron, 
as well as being chemically homogenous, with $<$[O/H]$>$=+0.35$ \pm$ 0.06 and $<$[Na/H]$>$=+0.51 $\pm$ 0.07.

We do not find evidence of two populations in sodium in this cluster.  
Although the sample observed here with APOGEE is relatively small, it does
cover the temperature range from 3500K (well up the luminous part of the RGB) to roughly 
4500K (for stars near the base of the RGB, as well as two stars in the RC). 
This sample is well-represented by a single Na abundance, with
a star-to-star abundance scatter consistent with the expected uncertainties in the analysis. 

The results of the non-LTE calculations indicate that the two Na I lines used in this
analysis ($\lambda$16373.86\AA\ and $\lambda$16388.85\AA) are largely free from departures 
from LTE over the T$_{\rm eff}$ and log g ranges covered by the high-metallicity 
red giants analyzed here (with all corrections being at most 0.04 dex) and so 
non-LTE corrections have no significant impact on our conclusions.

\acknowledgments
\section{Acknowledgments}

V.S., S.R.M., and J.H. acknowledge partial support from the National Science Foundation (AST1109888).
T.C.B. acknowledges partial support from PHY 08-22648; Physics Frontier Center/{}Joint Institute or Nuclear
Astrophysics (JINA), PHY 14-30152; Physics Frontier Center/{}JINA Center for the Evolution of the Elements (JINA-CEE), 
from the National Science Foundation.
D.G. acknowledges support from the Chilean BASAL Centro de Excelencia en Astrofisica
y Tecnologias Afines (CATA) grant PFB-06/2007.
S.M. acknowledges support from the Australian Research Council through DECRA Fellowship DE140100598.
This publication makes use of data products from the Two Micron All Sky Survey, which is a joint project of the
University of Massachusetts and the Infrared Processing and Analysis Center/California Institute of Technology, 
funded by the National Aeronautics and Space Administration and the National Science Foundation.  This research 
has made use of the SIMBAD database, operated at CDS, Strasbourg, France.

Funding for SDSS-III has been provided by the Alfred
P. Sloan Foundation, the Participating Institutions, the
National Science Foundation, and the U.S. Department
of Energy Office of Science. The SDSS-III web site is
http://www.sdss3.org/.

SDSS-III is managed by the Astrophysical Research
Consortium for the Participating Institutions of the
SDSS-III Collaboration including the University of Arizona, the Brazilian Participation Group, Brookhaven 
National Laboratory, University of Cambridge, Carnegie
Mellon University, University of Florida, the French
Participation Group, the German Participation Group,
Harvard University, the Instituto de Astrofisica de Canarias, the Michigan State/Notre Dame/JINA 
Participation Group, Johns Hopkins University, Lawrence Berkeley National Laboratory, 
Max Planck Institute for Astrophysics, New Mexico State University, New York University, 
The Ohio State University, Pennsylvania State University, University of Portsmouth, Princeton University, the
Spanish Participation Group, University of Tokyo, University of Utah, 
Vanderbilt University, University of Virginia, University of Washington, and Yale University.



{\it Facilities:} \facility{SDSS}

\clearpage

\clearpage

\clearpage
\begin{deluxetable}{ccccccccccc}
\tabletypesize{\scriptsize}
\tablecaption{Derived Stellar Parameters and Abundances}
\tablewidth{0pt}
\tablehead{
\colhead{2MASS ID} &
\colhead{SBG} &
\colhead{K$_{s}$$_{\rm 0}$} &
\colhead{(J-K$_{s}$)$_{0}$} &
\colhead{M$_{\rm bol}$} &
\colhead{T$_{\rm eff}$ (K)} &
\colhead{Log ${\rm g}$} &
\colhead{$\xi$ (Km s$^{-1}$)} &
\colhead{A(Fe)} &
\colhead{A(O)} &
\colhead{A(Na)}
}
\startdata
J19204557+3739509  & 5583 & 11.49  & 0.71  & 0.72  & 4500 & 2.45 & 1.3 & 7.89$ \pm $0.09  &  9.15$ \pm $0.06 &   6.75	$ \pm $0.05  	\\
J19204971+3743426  & 6963 & 7.77   & 1.16  & -2.41 & 3530 & 0.80 & 1.8 & 7.85$ \pm $0.09  &  9.02$ \pm $0.04 &   6.78	$ \pm $0.05	\\
J19205338+3748282  & 8266 & 9.72   & 0.89  & -0.76 & 4075 & 1.70 & 1.7 & 7.87$ \pm $0.16  &  9.11$ \pm $0.03 &   6.78	$ \pm $0.05	\\
J19205510+3747162  & 8904 & 9.54   & 0.93  & -0.89 & 4000 & 1.62 & 1.7 & 7.83$ \pm $0.12  &  9.13$ \pm $0.03 &   6.80	$ \pm $0.08	\\
J19205530+3743152  & 8988 & 11.04  & 0.79  & 0.42  & 4300 & 2.27 &  1.7 & 7.87$ \pm $0.04  &  9.15$ \pm $0.03 &   6.66	$ \pm $0.05	\\
J19210112+3742134  & 10898 & 10.89 & 0.81  & 0.31  & 4255 & 2.21 &  1.6 & 7.78$ \pm $0.09  &  9.10$ \pm $0.01 &   6.76	$ \pm $0.03	\\
J19210426+3747187  & 11814 & 10.05 & 0.84  & -0.50 & 4200 & 1.85 &  1.8 & 7.85$ \pm $0.12  &  9.14$ \pm $0.01 &   6.72	$ \pm $0.03	\\
J19210483+3741036  & 11957 & 11.51 & 0.68  & 0.68  & 4590 & 2.48 &  1.5 & 7.94$ \pm $0.08  &  9.20$ \pm $0.03 &   6.83	$ \pm $0.05	\\
J19211007+3750008  & 13260 & 11.92 & 0.74  & 1.24  & 4435 & 2.66 &  1.3 & 7.86$ \pm $0.08  &  9.07$ \pm $0.05 &   6.62	$ \pm $0.09	\\
J19211606+3746462  & 14379 & 7.66   & 1.13  & -2.54 & 3575 & 0.76 & 1.8 & 7.74$ \pm $0.16  &  9.03$ \pm $0.09 &   6.83	$ \pm $0.05	\\
J19213390+3750202  &  15966 & 8.75   & 1.02  & -1.59 & 3800 & 1.25 & 1.7 & 7.73$ \pm $0.06  &  9.02$ \pm $0.10 &   6.68	$ \pm $0.05	\\
\enddata

\tablecomments{SBG: Stetson et al. (2003)}
\end{deluxetable}

\clearpage





\begin{figure}
\epsscale{.80}
\plotone{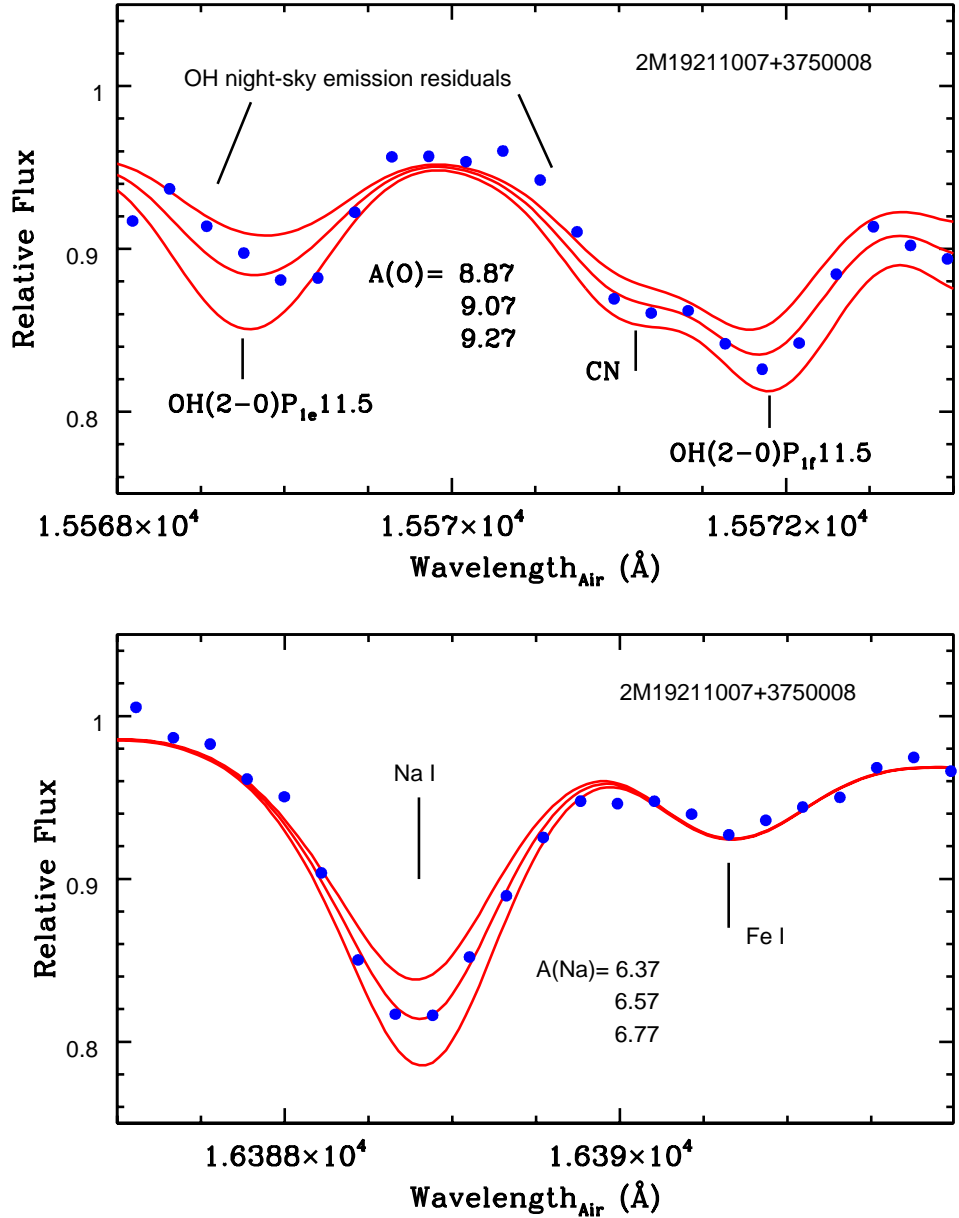}
\caption{Sample portions of APOGEE spectra (filled circles), along with synthetic spectra 
(solid curves) showing two of the
OH lines (top panel) and one of the Na I lines (bottom panel) in the NGC 6791 red giant
2M19211007+3750008 (T$_{\rm eff}$=4435K).  
\label{fig1}}
\end{figure}

\clearpage

\begin{figure}
\epsscale{.80}
\plotone{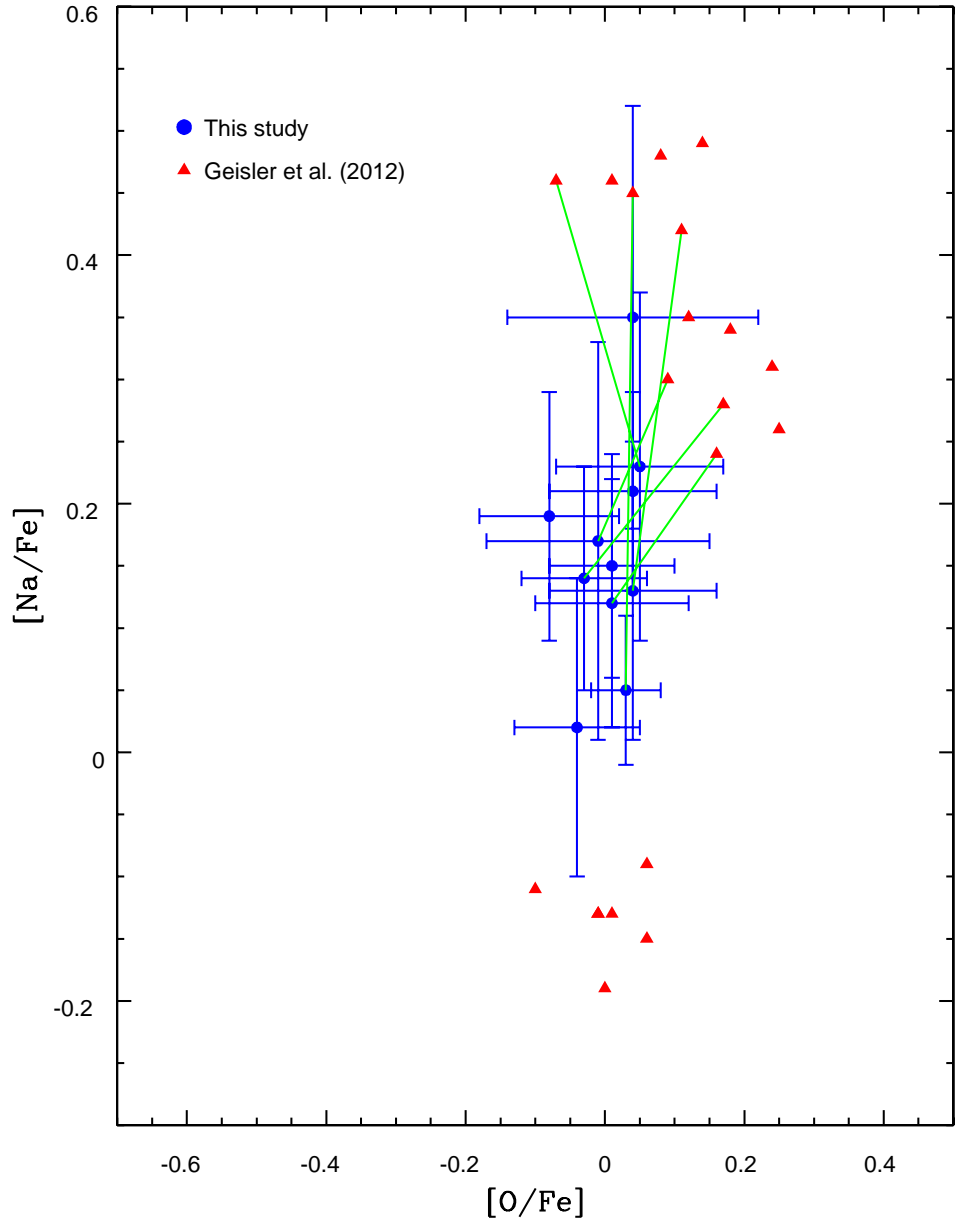}
\caption{Values of [Na/Fe] vs. [O/Fe] for NGC 6791. Blue circles are
results from this study, red triangles are from Geisler et al. (2012).
The solid green lines connect the abundance results for the six stars in common between 
this study and those of Geisler et al. (2012). 
\label{fig2}}
\end{figure}

\clearpage

\begin{figure}
\epsscale{.80}
\includegraphics[angle=-90,scale=0.60]{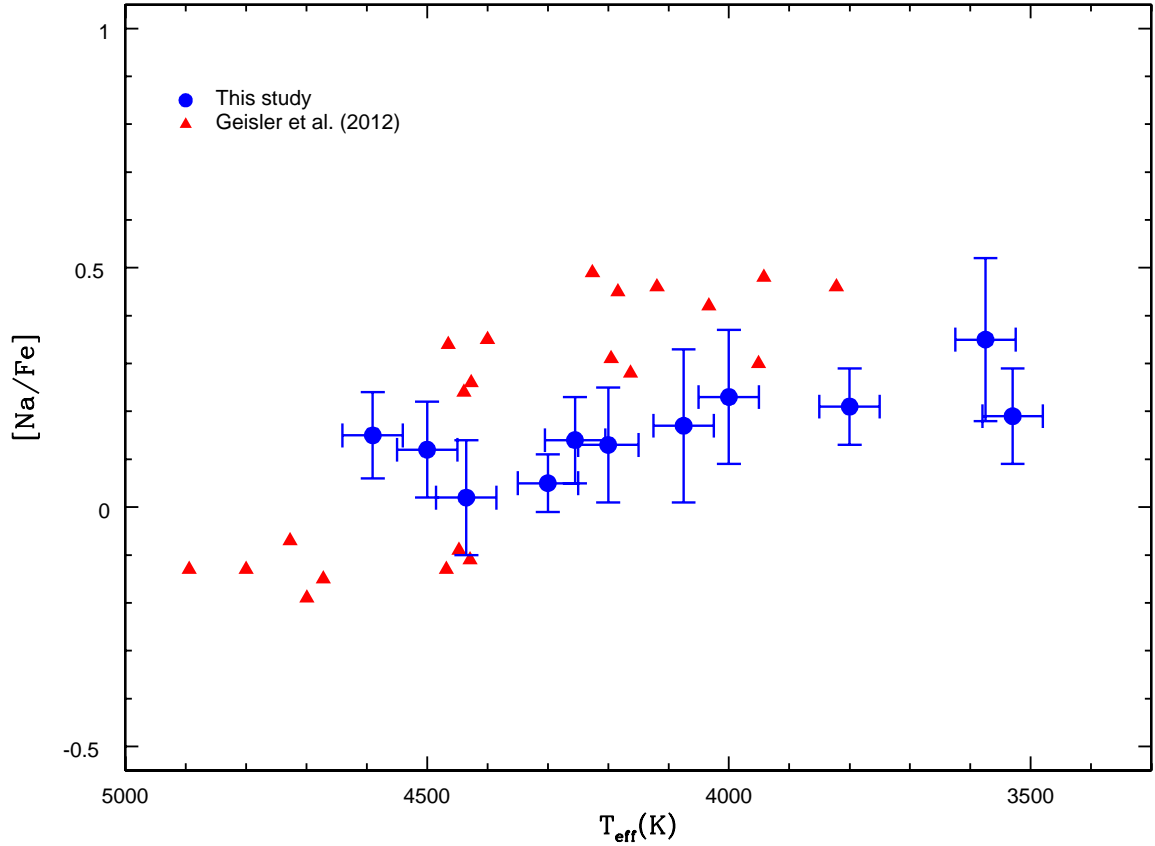}
\caption{[Na/Fe] vs. T$_{\rm eff}$ for NGC 6791 targets in this study (blue circles),
with error bars illustrating the estimated uncertainties. Results from Geisler et al. (2012; red triangles)
are also shown. There is no significant trend of APOGEE [Na/Fe] with effective temperature beyond the estimated
errors for this analysis, and the magnitude of departures from LTE are very small (less than 0.04 dex at most).
The Geisler et al. (2012) results indicate a significant change in [Na/Fe] values at T$_{\rm eff}$$\sim$4500K. 
\label{fig3}}
\end{figure}

\clearpage


\end{document}